\documentstyle[12pt,epsfig,amssymb]{article}
\def\ap#1#2#3{Ann.\ Phys.\ (NY) #1 (19#3) #2}
\def\cmp#1#2#3{Commun.\ Math.\ Phys.\ #1 (19#3) #2}

\def\np#1#2#3{Nucl.\ Phys.\ B#1 (19#3) #2}
\def\pl#1#2#3{Phys.\ Lett.\ #1B (19#3) #2}
\def\prb#1#2#3{Phys.\ Rev.\ B #1 (19#3) #2}
\def\prep#1#2#3{Phys.\ Rep.\ #1 (19#3) #2}

\def\frac#1#2{ {{#1} \over {#2} }}
\def\ie{\hbox{\it i.e.}{ }}

\def\half{\mbox{\small $\frac{1}{2}$}}
\def\re#1{(\ref{#1})}
\def\beq{\begin{equation}}
\def\eeq{\end{equation}}
\def\beeq{\begin{eqnarray}}
\def\beeqn{\begin{eqnarray*}}
\def\eeeq{\end{eqnarray}}
\def\eeeqn{\end{eqnarray*}}
\def\nome#1{{\label{#1}}}

\def\ltap{\raisebox{-.4ex}{\rlap{$\sim$}} \raisebox{.4ex}{$<$}}

\def\m{\mu}
\def\n{\nu}
\def\r{\rho}
\def\s{\sigma}

\def\G{\Gamma}
\def\eps{\epsilon}
\def\L{\Lambda}

\def\g{\gamma}
\def\D{\Delta}
\def\bc{\bar c}
\def\bchi{\bar \chi}
\def\bG{\bar\Gamma}
\def\de{\delta}
\def\t,#1{t^{#1}}
\def\f,#1#2#3{f^{#1 #2 #3}} 
\def\se{S_{\mbox{\footnotesize{eff}}}}

\def\Piinv{\Pi_{\mbox{\footnotesize{inv}}}}
\def\Pit{\tilde\Pi}
\def\DG{\D_{\G}}

\def\DGh{\hat{\D}_{\G}}
\def\LdL{\L\partial_\L}
\def\UV{\L_0\to\infty\;}
\def\p{\partial}
\def\K{K_{\L\L_0}}

\def\Kiu{K_{0\L_{0}}}
\def\Kin{K_{\L_{0}\infty}}

\def\dLID#1{ \frac {\L \partial D^{-1}_{\L\L_0}(#1)}{\partial \L}}

\def\d4#1{\frac {d^4 {#1} }{(2\pi)^4}}
\begin{document}
\begin{titlepage}
\begin{flushright}
UPRF-00-04 \\
\end{flushright}
\vspace{.4in}
\begin{center}
{\large{\bf Fine-tuning and the Wilson renormalization group}}
\bigskip \\ M. Bonini and E. Tricarico
\\
\vspace{\baselineskip}
{\small Universit\`a degli Studi di Parma \\ and\\
I.N.F.N., Gruppo collegato di Parma, 
\\viale delle Scienze, 
43100 Parma, Italy} \\
\mbox{} \\
\vspace{.5in}
{\bf Abstract} 
\bigskip 
\end{center} 
\setcounter{page}{0}
We use the Wilson renormalization group (RG) formulation to solve the
fine-tuning procedure needed in renormalization schemes breaking the
gauge symmetry.  To illustrate this method we systematically compute
the non-invariant couplings of the ultraviolet action
of the SU(2) pure Yang-Mills theory at one-loop order.



\end{titlepage}

\section{Introduction}

The Wilsonian renormalization group (RG) provides the most physical
framework to study general properties of quantum field theories. In
this formulation \cite{W,P} quantum fluctuations at short distances
are integrated out to give an effective action for longer distances.
In this approach the bare action can be though as the result of
integrating out the degrees of freedom with frequencies higher than the
ultraviolet (UV) scale $\L_0$ of the underlying theory (eventually
more fundamental). By further integrating out the fields with
frequencies up to some lower scale $\L$ one obtains the so called
Wilsonian effective action. In this process one needs to introduce a
scale which may conflict with symmetries.  For a gauge theory, the
gauge symmetry is completely broken at the intermediate scale $\L$ and
one has to show that the physical theory (\ie when all cutoffs are
removed) satisfies the Slavnov-Taylor (ST) identity.  This problem as
been investigated in refs.~\cite{B}-\cite{BV} where it has been
proved that by finely tuning the couplings in the Wilsonian effective
action the usual ST identity is recovered at $\L=0$, at least in
perturbation theory.  This fine-tuning provides some of the boundary
conditions of the RG flow.  In particular the breaking of the symmetry
can be studied at the physical point $\L=0$ \cite{noi} or at the UV
scale \cite{B,BV}.  In the former  case one enforces the relevant part
of the usual ST identity and determines the boundary conditions at the
physical point $\L=0$ of the relevant non-invariant couplings of the
Wilsonian effective action in terms of physical vertices evaluated at
the subtraction point.  Due to the non-locality of these vertices, the
fine-tuning is analytically difficult to solve (see ref.~\cite{noi}
where the boundary conditions of six relevant non-invariant couplings
of a pure gauge SU(2) theory at the first loop order were found in
this way).  In the latter the fine-tuning provides the couplings
of the bare action, which contains all possible relevant interactions.

The conflict between symmetries and regularization is a long-standing
problem in quantum field theory: in presence of chiral fermions no
consistent regularization is known to preserve chiral symmetry and all
possible non-invariant counterterms must be added to the classical
action in order to compensate the breaking of the symmetry generated
by the regularization. Although for non-anomalous theories this is an
algebraic solvable problem, the determination of these counterterms is
quite difficult in practice.  In this paper we will use the RG
formulation and the fine-tuning at the UV scale to provide all
non-invariant counterterms of the bare action at the one-loop order.
In order to keep things as simple as possible we will restrict to a
pure gauge $SU(2)$ theory. The inclusion of chiral fermions can be
done following the same lines \cite{chi,BV}.

The paper is organized as follows.  In section 2, after
introducing some notation, we give a brief description of the RG method
for the gauge case. In section 3 we recall the effective ST identity
for this theory. In section 4 we determine the non-invariant couplings
of the bare action at one-loop order by solving the fine-tuning at
$\L=\L_0$. Technical details of this calculation are given in the
appendix. In section 5 we present some concluding remarks and
hints for the generalization of this computation to higher
loops.

\section{Renormalization group flow}

Consider the $SU(2)$ gauge theory described by the classical
Lagrangian (in the Feynman gauge) 
$$
S_{YM}=\int d^4x \,
\left\{ -\frac 1 4 F_{\mu\nu} \,\cdot \, F^{\mu\nu} - \frac 1 2
\left(\p^\mu A_\mu \right)^2 - \bc\,\cdot\, \p^\mu D_\mu c \right\}\,,
$$
where the gauge stress tensor and the covariant derivative are
given by $F_{\mu\nu}=\p_\mu A_\nu -\p_\nu A_\mu + g \, A_\mu\,\wedge
\,A_\nu$, $D_\mu c =\p_\mu c + g \, A_\mu \,\wedge \, c$ and we have
introduced the usual scalar and external SU(2) products.  This action
is invariant under the BRS transformation \cite{BRS} 
$$ 
\de A^a_\mu=
\frac 1 g \eta \,D^{ab}_\mu c^b \,, \;\;\;\;\;\;\; 
\de c^a=-\half\,\eta\, \eps^{abc} c^b \,c^c \,, \;\;\;\;\;\;\; 
\de \bc^a= - \frac1 g \,\eta\, \p^\mu A^a_\mu  
$$ 
with $\eta$ a Grassmann parameter.

Introducing the sources $u^a_\mu$, $v^a$   associated to 
the variations of $A^a_\mu$, $c^a$ respectively 
one has the BRS action \cite{BRS}
\beq\nome{Stot}
S_{BRS}[\Phi,\g]
=
S_{YM} +
\int d^4x
\left\{
\frac 1 g u^a_\mu  D^{ab}_\mu c^b - \half \eps^{abc} v^a\,c^b \,c^c
\right\}
\eeq
where we have denoted by $\Phi$ and $\g$ the fields and the BRS 
sources
$$
\Phi_I=\{\,A^a_\mu, \,c^a, \,\bc^a\}
\,,
\;\;\;\;\;\;\;\;
\g_i=\{w_\mu^a,\, v^a\}
\,,
$$
and  $w_\mu^a=u_\mu^a+g\partial_\mu \bc^a$ (no source is introduced for 
$\bc^a$).

According to Wilson \cite{W,P} the
fields with frequencies $\L^2<p^2<\L_0^2$ 
are integrated in the path integral giving
\beq\nome{Z}
Z[J,\g]=N[J;\L,\L_0] \;\int {\cal D}\Phi \,
\exp i\{-
\half (\Phi\, D^{-1}\Phi)_{0\L}+(J\Phi)_{0\L}+\se\,[\Phi,\g;\L,\L_0]
\; \}
\,,
\eeq
where $N[J;\L,\L_0]$ contributes to the quadratic part of $Z[J,\g]$
and the functional $\se$ is the Wilsonian effective action which is the
generator of the connected amputated cutoff Green functions (except
the tree-level two-point function). $\se$ contains cutoff propagators
$D_{\L\L_0}$ given by the free propagators $D(p)$  multiplied
by a cutoff function $\K$ which is one for 
$\L^2\,\ltap\; p^2 \,\ltap\; \L_0^2$ and
rapidly vanishes outside. In \re{Z} we have also
introduced the cutoff scalar products between fields and
sources
$$
\half (\Phi\, D^{-1}\Phi)_{\L\L_0}
\equiv
\int_p\, K^{-1}_{\L\L_0}(p)\,p^2\;
\left\{\half A^a_\mu(-p)\,A^a_\mu(p)
- \bc^a(-p)\, c^a(p)\right\}\,,
\;\;\;\;\;\;
\int_p \equiv \int \frac{d^4p}{(2\pi)^4}\,,
$$
$$
(J\Phi)_{\L\L_0}\equiv
\int_p \, K^{-1}_{\L\L_0}(p) \;
\left\{j^a_\mu(-p)\, A^a_\mu(p) +
[\bchi^a(-p) - \frac i g p_\mu u^a_\mu(-p)] \, c^a(p) +
\bc^a(-p)\, \chi^a(p) \right\}
\,.
$$
Notice that at $\L=\L_0$ the Wilsonian effective action coincides 
with the bare action. 

The requirement that $Z[J,\g]$ is independent of the infrared cutoff $\L$
can be translated into a flow equation for $\se$, referred to as the
exact RG equation. For our purposes it is convenient to perform a
Legendre transform on $\se$ in order to obtain the so-called ``cutoff
effective action'' $\G[\Phi,\g;\L,\L_0]$, which is the generator of 1PI
cutoff vertex functions and reduces to the physical quantum effective
action in the limits $\L\to 0$ and $\UV$.  The flow equation can be 
written as \cite{Wett}-\cite{Mo}
\beq\nome{eveq}
\LdL \,\Pi [\Phi,\g;\L,\L_0]
=-\frac i2\int_q
\biggl[\dLID {q}\biggr]_{AB}\, (-1)^{\de_A}
\biggl[\frac{\de^2\G[\Phi,\g;\L,\L_0]}{\de\Phi_A(q)\de\Phi_B(-q)}\biggr]^{-1}
\,,
\eeq
where $\de_A=1$ if $\Phi_A$ is a fermionic field and $0$ otherwise and 
\beq\nome{pigreco}
\Pi[\Phi,\g;\L,\L_0]=\G[\Phi,\g;\L,\L_0]+\half (\Phi,\,D^{-1}\Phi)_{\L\L_0}-
\half (\Phi,\,D^{-1}\Phi)_{0\L_0}\,.
\eeq
This equation can be integrated by giving boundary conditions in $\L$.
In order to set them we distinguish between relevant couplings and
irrelevant vertices according to their mass dimension.  Relevant
couplings have non-negative mass dimension and are defined as the
value of some vertices and their derivatives at a given normalization
point (see \cite{noi} for their definition).
For the  $SU(2)$ theory there are nine relevant couplings 
which are the coefficients of the following 
monomials
\beeq\nome{mono}&&
\half A_\mu (g_{\mu\nu}\partial^2-\partial_\mu\partial_\nu)   
\cdot A_\nu\,,
\;\;\;\;\;
w_\mu\cdot \partial_\mu \,c\,,
\;\;\;\;\;
\half  A_\mu \cdot A_\mu\,,
\;\;\;\;
\half (\partial_\mu A_\mu)^2\,,
\nonumber\\&&
\nonumber\\&&
(\partial_\nu A_\mu) \cdot A_\mu \wedge A_\nu\,,
\;\;\;\;
w_\mu \cdot c \wedge A_\mu \,,
\;\;\;\;
\half v\cdot c \wedge c\,,
\\&&
\nonumber\\&&
\frac 14 (A_\mu\wedge A_\nu)\cdot(A_\mu\wedge A_\nu)
\,,\;\;\;\;
\frac 14 (A_\mu\cdot A_\nu) \cdot (A_\mu\cdot A_\nu)
\,.\nonumber
\eeeq
To set the boundary conditions, one observes that the cutoff effective
action $\G[\Phi,\g;\L,\L_0]$ at $\L=\L_0$ becomes the bare action and
to ensure perturbative renormalizability it must be local, thus
irrelevant vertices must vanish at $\L=\L_0$. 
As far as the relevant couplings, the boundary conditions must provide the
physical coupling $g(\m)$ and guarantee symmetry.
In fact the cutoff $\L$ explicitly breaks gauge invariance and one has
to show that at the physical point $\L=0$ and $\UV$ the
Slavnov-Taylor identity can be recovered.
As shown in refs.~\cite{B,noi} the symmetry is
ensured if the relevant couplings are properly fixed at some value of $\L$. 
In the next section we briefly recall how this procedure can be implemented.

Once the boundary conditions are given, equation \re{eveq}
can be thought as an alternative definition of the theory which in
principle is non-perturbative. 
On the other hand the iterative solution of \re{eveq} reproduces 
the usual loop expansion and, as shown by Polchinski~\cite{P},  
the $\UV$ limit of $\G[\Phi,\g;\L,\L_0]$ is finite for all values of 
$\L\,$~\footnote{For a generalization of this proof to gauge theories see 
for instance \cite{noi}.}.  
In the following this limit will be taken and $\G[\Phi,\g;\L,\infty]$
will be denoted by $\G[\Phi,\g;\L]$. Notice that, once this limit has
been performed, the irrelevant vertices of $\G[\Phi,\g;\L]$ do not
vanish at $\L=\L_0$ but are only suppressed by inverse powers of
$\L_0$.

\section{Effective ST identity} 
In the RG formulation the ST identity for the physical effective action 
$\G[\Phi,\g]\equiv\G[\Phi,\g;0,\infty]$ is ensured if the cutoff 
effective action satisfies a modified ST \cite{B}-\cite{BV}.
As usual in studying ST identities, it is convenient to remove from 
the functional $\Pi$ given in \re{pigreco} the gauge fixing term 
and introduce the Slavnov operator \cite{Becchi0}
$$
{\cal S}_{\Pi'}= -i \int_p \left( \frac{\de \, \Pi'}{\de \Phi_i(-p)}
\frac{\de\, }{\de \g_i(p)} + \frac{\de \, \Pi'}{\de \g_i(p)} 
\frac{\de\, }{\de \Phi_i(-p)}\right)\,, \;\;\;\; \Phi_i=\{A_\m,\,c \} 
$$
with 
$$
\Pi'[\Phi,\g;\L]=\Pi[\Phi,\g;\L]
+\half\int_p p_\m\, p_\n \,A_\m(-p)\cdot A_\n(p).
$$ 
The modified ST identity then reads 
\beq\nome{STeff}
\DG(\L)\equiv
{\cal S}_{\Pi'} \Pi'[\Phi,\g;\L]+\DGh(\L)=0\, ,
\eeq
where 
\beeq\nome{dgh}
&&\DGh[\Phi,\g;\L]
=-\hbar\,
\int_{p,q} K_{0\L}(p)
(-1)^{\de_L}\left[D^{-1}_{\L \infty} (-p)\right]_{Li} \, 
\left[\frac{\de^2\G[\Phi,\g;\L]}{\de\Phi_L(-p)\de\Phi_J(-q)}\right]^{-1}
\nonumber\\&&
\phantom{\DGh[\Phi,\g;\L]=-\hbar\,\int_{p,q}}
\times 
\, \frac{\de^2}{\de\Phi_J(q)\de\g_i(p)}
\left(\Pi[\Phi,\g;\L]-\frac1g \int_{x}u_{\m}\p_{\m}c\right)
\,.
\eeeq
The factor $\hbar$ has been inserted to put into evidence that $\DGh$
vanishes at tree-level.  Notice that at the physical point $\L=0$ the
gauge symmetry condition \re{STeff} reduces to the usual ST identity,
since $\Pi$ becomes the physical effective action and $\DGh$ vanishes.
It can be shown that if $\DG$ vanishes up to loop ${\ell}$ then at the
next loop $\DG$ is $\L$-independent and therefore can be analyzed at
every value of $\L$. If one imposes \re{STeff} at $\L=0$, one gets the
boundary conditions for the relevant part of the effective action
given in terms of the physical coupling $g(\mu)$.
Alternatively, in \re{STeff} one can choose $\L$ much bigger than all
external momenta, namely $\L=\L_0$, and determine the cutoff dependent
couplings of UV action.  Both these procedures provide the boundary
conditions of the RG flow for the non-invariant relevant couplings of
the cutoff effective action.  We will adopt the latter possibility in
the present paper.

At $\L=\L_0$ the functional $\DG$ is local, or more precisely, its
irrelevant contributions vanish as inverse powers of $\L_0$ and
disappear in the $\UV$ limit (see \cite{MT} for a proof).  This is
obviously true for the functional ${\cal S}_{\Pi'}\, \Pi'(\L_0)$, due
to the boundary conditions imposed on the irrelevant vertices of
$\Pi$, and can be shown by a direct calculation for the functional
$\DGh(\L_0)$.  Therefore in the $\UV$ limit the effective ST identity
\re{STeff} reduces to a finite set of equations 
(the so called fine-tuning equations)
obtained by requiring the vanishing of relevant part of $\DG$.
In fact this system of equations overdetermines the UV couplings and 
in order to prove its solvability one must exploit  the so-called 
consistency conditions which reduce the number of independent equations
\cite{B}.
In the next section we will solve these fine-tuning equations
at one-loop order tuning the couplings of the UV 
action at this order. 
Besides determining these couplings  this calculation 
provides a direct check of the solvability of the fine-tuning problem.

\section{One-loop computations}
At one-loop and at
$\L=\L_0$, the fine-tuning equation \re{STeff} reduces to 
\beq \nome{fintun}
{\cal S}_{\Pi^{(0)}}\, \Pi^{(1)}(\L_0)\,=-\, \DGh^{(1)}(\L_0)\,,
\eeq
where the local functional 
$\Pi^{(1)}(\L_0)\,\equiv\,\Pi^{(1)}[\Phi,\g;\L_0]$ contains the
relevant monomials given in \re{mono} 
(as discussed in the previous section its irrelevant contributions 
vanish as $\UV$ and therefore can be omitted in the calculations 
performed in this section).
This functional can be split
into two contributions~\footnote{For the sake of simplicity the loop
order index will be sometimes  understood.}
\beq \nome{pigreco1}
\Pi(\L_0)=\Piinv(\L_0)+\Pit(\L_0)\, ,
\eeq
where $\Piinv$ satisfies the ST identity 
\ie ${\cal S}_{\Piinv}\, \Piinv=0$. The explicit
form of $\Piinv$ is 
$$
\Piinv(\L_0)=\int \mbox{d}^4 x\, \Bigg\{-\frac14\,z_1
\, 
{\cal F}_{\m\n}\,\cdot\, 
{\cal F}^{\m\n} + z_2\,z_3 \,\left( \frac1{g\, z_3}\, w_\m\,\cdot\, 
{\cal D}_\m c
-\frac12\, v\,\cdot\,c\,\wedge\,c \right) 
\Bigg\}\, ,
$$
with ${\cal F}_{\mu\nu}=\p_\mu A_\nu -\p_\nu A_\mu
+ g \, z_3 \, A_\mu \,\wedge\,A_\nu$ and the covariant derivative
given by
${\cal D}_\mu c =\p_\mu c + g \, z_3 \,A_\mu \,\wedge\, c$. 
The remaining monomials contribute to $\Pit$ which can be written as
\beeq \nome{pitilde}
\Pit(\L_0)\!\!\!& \equiv&\!\!\! \int\mbox{d}^4x\,
\bigg\{
\s_1 \,\L_0^2\, \frac 1 2 A_\mu^2
+\s_2 \,\frac 1 2 (\partial_\mu A_\mu)^2
+\s_3\, w_\mu \cdot c\wedge A_\mu
+\s_4 \, \frac 1 2 v\cdot c \wedge c
\nonumber\\
&+&\!\!\!
\s_5\, \frac{g^2}{4}(A_\mu\wedge A_\nu)^2
+\s_6\, \frac{g^2}{4} (A_\mu\cdot A_\nu)^2
\bigg\}\,.
\eeeq
The functional $\Pi(\L_0)$ is equal to the UV action $S_{\L_0}[\Phi,\g]$
with  the UV wave function constants and UV couplings given by
$$
z_i=z_i(g,\mu/\L_0)
\,,\;\;\;\;
\s_i=\s_i(g,\mu/\L_0)
\,.
$$ 
Notice that the fine-tuning equation \re{fintun} allows to compute the
couplings in $\Pi^{(1)}(\L_0)$ since $\DGh^{(1)}(\L_0)$ depends only
on $\Pi^{(0)}=S_{BRS}$.  As a matter of fact only the couplings $\s_i$ are
tuned with this procedure.  On the contrary the couplings $z_i$ are
not involved in the fine-tuning.  They are free parameters which can
be fixed setting the field normalization and the gauge coupling at the
subtraction point $\m$ equal to their physical values at $\L=0$, \ie
$z_i(\L=0)=1$.  In the standard language this corresponds to the
renormalization prescriptions.  In the following subsections we will
determine the expression for the couplings $\s_i$ in term of the
Yang-Mills coupling $g$ and the cutoff function.  The values of these
couplings corresponding to three different choices of the cutoff
function are collected in the tables~$1$-$3$ given below.

Using the parameterization given in \re{pigreco1} and \re{pitilde}, the
l.h.s. of \re{fintun} becomes
\beeq 
&&{\cal S}_{\Pi^{(0)}}\, \Pi^{(1)}(\L_0)\,=\int d^4x \, 
\bigg\{
\de_1\L_0^2 \,A_\mu\cdot\partial_\mu c
+\de_2 \,A_\mu\cdot\partial^2\partial_\mu c
+\de_3 \,A_\mu\cdot (\partial^2 A_\mu)\wedge c
\bigg.
\nonumber\\
&&\phantom{{\cal S}_{\Pi^{(0)}}}
\bigg.
+\de_4 \,A_\mu\cdot (\partial_\mu \partial_\nu A_\nu)\wedge c
+\half \de_5 \,(\partial_\mu w_\mu)\cdot c\wedge c
+\half \de_6 \,(w_\mu \wedge A_\mu) \cdot (c\wedge c)
\bigg.
\nonumber\\
&&\phantom{{\cal S}_{\Pi^{(0)}}}
\bigg.
+\de_7 \, ((\partial_\mu A_\mu) \cdot A_\nu) (A_\nu \cdot c)
+\de_8 \, ((\partial_\mu A_\mu) \cdot c) (A_\nu \cdot A_\nu)
+\de_9 \, ((\partial_\nu A_\mu) \cdot A_\nu) (A_\mu \cdot c)
\bigg.
\nonumber\\
&&\phantom{{\cal S}_{\Pi^{(0)}}}
\bigg.
+\de_{10} \, ((\partial_\nu A_\mu) \cdot A_\mu) (A_\nu \cdot c)
+\de_{11} \, ((\partial_\nu A_\mu) \cdot c) (A_\mu \cdot A_\nu)
\bigg\} 
\nonumber
\eeeq
where 
\beeq
\label{de}
&&
\de_1=-\frac i g \s_1\,,
\;\;\;\;\;\;\;\;\;\;
\de_2=\frac i g \s_2\,,
\nonumber\\
&&
\de_3=-i\s_3\,,
\;\;\;\;\;\;\;\;\;\;
\de_4=-i(\s_2-\s_3)\,,
\nonumber\\
&&
\de_5=\,\frac1g\,\de_6=\frac{i}{g}(\s_3-\s_4)\,,
\nonumber\\
&&
\de_7=\de_9=\de_{11}=i g (\s_3+\s_6-\s_5)\,,
\;\;\;\;\;\;\;\;\;\;
\de_8=\frac12\de_{10}=i g (\s_5-\s_3)\,.
\eeeq
The system of equations obtained by imposing $\de_i$ to be equal to the
coefficient of the corresponding monomials in $-\DGh$
overdetermines the couplings $\s_i$ (eleven equations and  five
unknown couplings) therefore it can be solved if one has only five 
independent equations.
The necessary  six relations among the parameters $\de_i$
are automatically obtained from the anticommutativity of the BRS
operator ${\cal S}_\Pi$, as can be seen in \re{de}. 
Though it is not evident that the same relations hold for the
coefficients of $\DGh$, the direct calculation given in the following
subsections will shown that this is true.

At one-loop oder $\DGh$ can be evaluated taking the integrand in
\re{dgh} at tree level and noticing that $\Pi^{(0)}=S_{BRS}$ and then 
the sum over $\Phi_J$  is restricted to the $(A,c)-$fields. 
At $\L=\L_0$,  $\DGh^{(1)}$ is given by
\beq\nome{dgh1}
\DGh^{(1)}(\L_0)
=\int_{p,q} K_{0\L_0}(p)(-)^{\de_i}\bigg(D_{\L_0\infty}(q)\,
\bG^{(0)}[-q,-p;\L_0]\bigg)_{ji} 
\,\, \frac{\de^2}{\de\Phi_j(q)\de\g_i(p)}
\left(S_{BRS}-\frac1g \int_{x}u_{\m}\p_{\m}c\right)
\,,
\eeq
where 
the functional $\bG^{(0)}$ originating from the 
inversion of $\frac{\de^2\G[\Phi,\g;\L]}{\de\Phi(-q)\de\Phi(-p)}$
is recursively defined as 
\beeq\nome{bG0}
&&\bG^{(0)}_{IJ}[-q,-p;\Phi,\g;\L_0]=
(-)^{\de_J}
\bigg(S_{BRS}^{int}(-q,-p)\bigg)_{IJ}
\nonumber\\
&&\phantom{\bG^{(0)}_{IJ}[-q,-p;}
-\int_{q'} \bigg(D_{\L_0\infty}(q')\,\bG^{(0)}[-q',-p;\Phi,\g;\L_0]\bigg)_{LJ}
\bigg(S_{BRS}^{int}(-q,q')\bigg)_{IL}
\,,
\eeeq 
with $S_{BRS}^{int}$ the interaction part of $S_{BRS}$ and
$$
\bigg(S_{BRS}^{int}(p,q)\bigg)_{IJ}=
\frac{\delta}{\delta \Phi_J(q)}\frac{\delta}{\delta \Phi_I(p)}S_{BRS}^{int}\,.
$$ 
For instance when $\Phi_I$ and $\Phi_J$ are vector fields one has
\beq\nome{ex}
\bigg(S_{BRS}^{int}(p,q)\bigg)_{A_\n^b\,A_\m^a}=
-ig\eps^{abc}\,t_{\m\n\r}(p,q)\,A^c_\r(-p-q)
-g^2\, t_{\m\n\r\s}^{abcd}\,\int_k A_\r^c(k)\,A_\s^d(-p-q-k)\,,
\eeq
where
$$
t_{\m\n\r}(p,q)=(p-q)_\r g_{\m\n}+(2q+p)_\m g_{\n\r}-(2p+q)_\n g_{\m\r}
$$ 
and 
\beeq&&
t^{a_1 \cdots a_4}_{\mu_1 \cdots \mu_4}
=
\left(\eps^{a_1a_2c}\eps^{ca_3a_4}-\eps^{a_1a_4c}\eps^{ca_2a_3}
\right)g_{\mu_1\mu_3}g_{\mu_2\mu_4} 
+
\left(\eps^{a_1a_3c}\eps^{ca_2a_4}-\eps^{a_1a_4c}\eps^{ca_3a_2}
\right)g_{\mu_1\mu_2}g_{\mu_3\mu_4} 
\nonumber\\&&
\nonumber\\&&
\phantom{t^{a_1 \cdots a_4}_{\mu_1 \cdots \mu_4}}
+
\left(\eps^{a_1a_3c}\eps^{ca_4a_2}-\eps^{a_1a_2c}\eps^{ca_3a_4}
\right)g_{\mu_1\mu_4}g_{\mu_2\mu_3}
\nonumber
\eeeq
are the three and four-vector elementary vertices, respectively.

In the following subsections we compute the coefficients of the
various field monomials in $\DGh$.

\subsection{$Ac$-vertex}
There are three contributions~\footnote{This computation can be found
in \cite{BV} and is repeated here for completeness.} to 
the $A_\m\cdot c-$vertex which are obtained inserting in \re{dgh1}
the first term of \re{bG0} and setting $(\g_i=w_\n\,, \Phi_j=A_\n)$,
$(\g_i=w_\n\,, \Phi_j=c)$ and $(\g_i=v\,, \Phi_j=c)$.  The corresponding
graphs are given in Fig~1. 
\begin{figure}[htbp]
\begin{center}
\epsfysize=4cm
\epsfbox{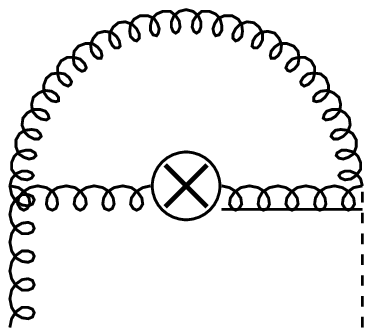}
\epsfysize=4cm
\epsfbox{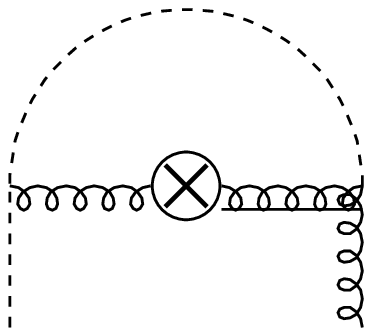}
\epsfysize=4cm
\epsfbox{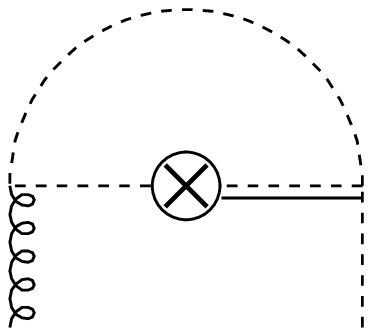}
\end{center}
\caption{\small{Graphical contribution to the $A$-$c$-vertex of $\DGh$.
The curly and 
dashed line denotes the gluon and ghost field 
respectively; the double lines represent the BRS source
associated to the field depicted by the top line. All momenta are incoming.
}}
\end{figure}
\newline
Using the vertices of $S_{BRS}$ one obtains
$$
\int_p A_\m(p)\cdot c(-p) 
\hat{\D}_{\G\,\m}^{(Ac)}(p;\L_0)
$$
with
$$
\hat{\D}_{\G\,\m}^{(Ac)}(p;\L_0)\,=\,
-2ig\int_q \bigg(t_{\n\m\n}(q,p)-2q_\m\bigg)
\frac{\Kin(q)}{q^2}\Kiu(q+p)\,.
$$
The presence of the two cutoff functions having
almost non-intersecting supports
(\ie $q^2 \gtrsim \L_0^2$, $(q+p)^2 \lesssim \L_0^2$) 
ensures that this vertex is quadratically divergent in the $\UV$ limit 
but without logarithmic divergences. 
In this limit one gets
$$
\hat{\D}_{\G\,\m}^{(Ac)}(p;\L_0)\,
= i\,p_\m\,[\hat\de_1\,\L_0^2-\hat\de_2\,p^2
\,+\, {\cal O}(p^4/\L_0^2)]
$$
with
\beq\nome{dh1}
\hat\de_1=\frac{2g}{\L_0^2}\int_q\frac{(1-K)(3K+2K' q^2)}{q^2}\,,
\eeq
and 
\beq\nome{dh2}
\hat\de_2=-2g\int_q\frac{(1-K)(18K'+21K''q^2+4K''' q^4)}{6 q^2}\,,
\eeq
where $K\equiv\Kiu(q)$ and the $'$ denotes the derivative with respect
to $q^2$ and we have used the identity $\Kin(q)=1-\Kiu(q)$. 
The coefficients $\hat\de_1$ and $\hat\de_2$ are 
finite and  cutoff function dependent.
Recalling that $\Kiu(0)=1$ and that $\Kiu(q)\to 0$
rapidly enough for $q^2\to \infty$, \re{dh1} and \re{dh2} 
can be rewritten as 
\beq\nome{dh1'}
\hat\de_1=\frac{2g}{\L_0^2}\int_q\frac{(1-2K)K}{q^2}
\eeq
and
\beq\nome{dh2'}
\hat\de_2=g\,\biggl(\frac{5r}{6}-3\int_qK'^2\biggr)\,,
\eeq
where $r=i/(16\pi^2)$ (the factor $i$ comes from the
$q$-integration).  The fine-tuning equation \re{fintun} for the
$Ac$-vertex becomes $$
\de_1=-\hat\de_1\,, \;\;\;\;\;\;
\de_2=-\hat\de_2\,.
$$
Then from \re{de} one finds that $\s_1$ and $\s_2$ are given by 
\beq\nome{sig1}
\s_1^{(1)}=-\frac{2ig^2}{\L_0^2}\int_q\frac{(1-2K)K}{q^2}\,,
\eeq
\beq\nome{sig2}
\s_2^{(1)}=ig^2\,\biggl(\frac{5r}{6}-3\int_qK'^2\biggr)
\eeq
and their values depend, as shown in table~1,  on the cutoff function.
\begin{table}[htbp]
\begin{center}
\begin{tabular}{|c |c | c| c |} \hline
                   &   &  &      \\
$\Kiu(p)$ & 
$e^{-p^2/\L_0^2}$&
$\frac{\L_0^4}{(p^2+\L_0^2)^2} $& 
$\frac{\L_0^{2n}}{p^{2n}+\L_0^{2n}} $
\\
 & & & \\  \hline
 & & &  \\
$\s_1$ &  
0&
-$\frac{2ig^2r}3$ &
$\frac{2\pi ig^2 r(n-2)}{n^2sin(\pi/n)}$ 
\\
 & & & \\  \hline
 & & &  \\
$\s_2$ &  
$\frac{ig^2r}{12}$&
$ \frac{7ig^2r}{30}$ &  
$ \frac{i g^2 r(5-3n)}{6}$ 
\\
  &  &  &   \\
\hline
\end{tabular} 
\end{center}
 \caption[Table 1] {\small{One-loop values of the couplings $\s_1$ and $\s_2$ 
for three different cutoff functions ($r=\frac{i}{16\pi^2}$ and $n\ge 1$).}}
\end{table}

\subsection{$AAc$-vertex}
In order to find the $AAc$-vertex one has to insert in \re{dgh1} the
functional $\bG^{(0)}$ obtained from the first iteration in \re{bG0}, \ie
\beq\nome{bG1}
- (-)^{\de_J}\int_{q'} \bigg(D_{\L_0\infty}(q')\,
S_{BRS}^{int}(-q',-p)\bigg)_{LJ}
\bigg(S_{BRS}^{int}(-q,q')\bigg)_{IL}
\,.
\eeq
One finds one contribution for $(\g_i=w_\n\,, \Phi_j=A_\n)$,
two for $(\g_i=w_\n\,, \Phi_j=c)$ 
and one for $(\g_i=v\,, \Phi_j=c)$
(plus the terms obtained from the permutation of the two vectors).
\begin{figure}[htbp]
\begin{center}
\epsfysize=4cm
\epsfbox{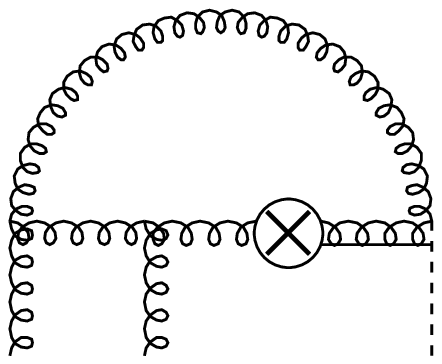}
\epsfysize=4cm
\epsfbox{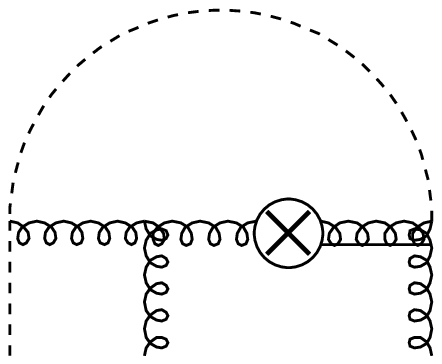}
\epsfysize=4cm
\epsfbox{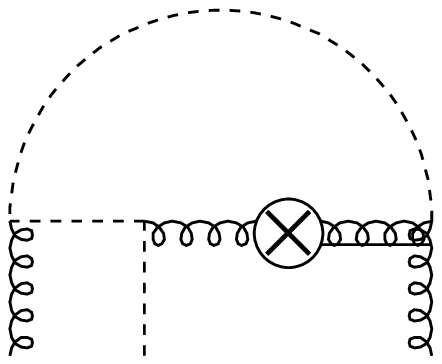}
\epsfysize=4cm
\epsfbox{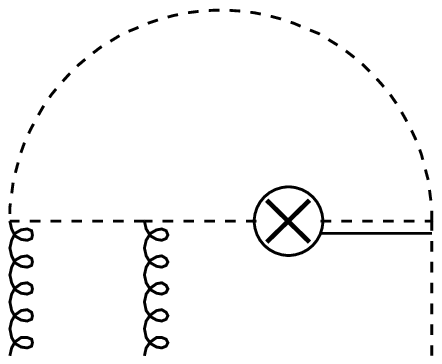}
\end{center}
\caption{\small{Graphical contribution to the $A$-$A$-$c$-vertex of $\DGh$.
}}
\end{figure}
The corresponding
graphs are given in Fig~2. 
Using the vertices of $S_{BRS}$ one obtains
$$
\frac12\int_{p_1p_2} A_{\m_1}(p_1)
\cdot A_{\m_2}(p_2)\wedge c(-p_1-p_2) \;
\hat{\D}_{\G\,\m_1\m_2}^{(AAc)}(p_1,p_2;\L_0)
$$
with
\beeq
&&\hat{\D}_{\G\,\m_1\m_2}^{(AAc)}(p_1,p_2;\L_0)\,
\nonumber\\&&
\;\;\;\;
=-\frac{g^2}{2}\,\int_q\Biggl[
t_{\r\m_1\n}(q,p_1)\,t_{\n\m_2\r}(q+p_1,p_2)
\frac{\Kin(q)\Kin(q+p_1)}{q^2 (q+p_1)^2}
\Kiu(q+p_1+p_2)
\nonumber\\&&
\nonumber\\&&
\;\;\;\;
-(q+p_1+p_2)_\n t_{\n\m_1\m_2}(q,p_1)
\frac{\Kin(q)\Kin(q+p_1+p_2)}{q^2 (q+p_1+p_2)^2}
\Kiu(q+p_1)
\nonumber\\&&
\nonumber\\&&
\;\;\;\;
-q_{\m_1}(q+p_1)_{\m_2}
\frac{\Kin(q)\Kin(q+p_1)}{q^2 (q+p_1)^2}
(\Kiu(q-p_2)+\Kiu(q+p_1+p_2))
\nonumber\\&&
\nonumber\\&&
\;\;\;\;
-(1\leftrightarrow 2)
\Biggr]\,.
\nonumber 
\eeeq
In the $\UV$ limit one gets 
$$
\hat{\D}_{\G\,\m_1\m_2}^{(AAc)}(p_1,p_2;\L_0)\,
= \,g_{\m_1\m_2}(p_1^2-p_2^2)\,\hat\de_3\,+(p_{1\m_1}\,p_{1\m_2}-
p_{2\m_1}\,p_{2\m_2})\hat\de_4\,
\,+\, {\cal O}(P^4/\L_0^2)\,,
$$
where $P$ is a combination of the external momenta $p_1$ and $p_2$,
\beq\nome{dh3}
\hat\de_3=-g^2\int_q\frac{(1-K)}{6q^4}[13K^2-2q^2(11K'+8K'^2q^2+K''q^2)+
K(-13+15K'q^2-8K''q^4)]
\eeq
and 
\beq\nome{dh4}
\hat\de_4
=g^2\int_q\frac{(1-K)}{6q^4}[13K^2+2q^2(-8K'+10K'^2q^2-K''q^2)+
K(-13+45K'q^2+10K''q^4)]\,.
\eeq
The fine-tuning equation \re{fintun} for the $AAc$-vertex 
allows to find the value of $\s_3$ in term of $g$
and the cutoff function. From $\de_3=\hat\de_3$ and \re{de} one finds
\beq\nome{sig3}
\s_3^{(1)}=-ig^2\,\biggl(\frac{37r}{36}-
\int_q\frac{13K(1-K)^2+10q^4 K'^2}{6q^4}\biggr)
\eeq
and its value for different choices of the cutoff function is given in table~2.
\begin{table}[htbp]
\begin{center}
\begin{tabular}{|c |c | c| c |} \hline
                   &   &  &      \\
$\Kiu(p)$ & 
$e^{-p^2/\L_0^2}$&
$\frac{\L_0^4}{(p^2+\L_0^2)^2} $& 
$\frac{\L_0^{2n}}{p^{2n}+\L_0^{2n}} $
\\
 & & & \\  \hline
 & & &  \\
$\s_3$ &  
-$\frac{ig^2r(39\ln\frac43+11)}{18}$&
 $\frac{49igr}{360}$&
-$\frac{igr(10n^2-37n+39)}{36n}$
\\
  &  &  &   \\
\hline
\end{tabular} 
\end{center}
 \caption[Table 2]{\small {One-loop values of the coupling $\s_3$ for three
 different cutoff functions ($r=\frac{i}{16\pi^2}$ and $n\ge 1$).}}
\end{table}

In order to check the consistency of our computation the fine-tuning
equation $\de_4\equiv -i(\s_2-\s_3)=\hat\de_4$ must be automatically 
satisfied with the couplings $\s_2$ and $\s_3$ previously determined and
given in \re{sig2} and \re{sig3}, respectively. Alternatively one has
to prove the consistency condition $\hat\de_3+\hat\de_4=-g\hat\de_2$. 
From \re{dh3} and \re{dh4} one finds
$$
-\frac1g(\hat\de_4+\hat\de_3)=
-g\int_q\frac{1-K}{q^2}(K'+5KK'+6K'^2q^2+3KK''q^2)
$$
which becomes equal to \re{dh2'} after integration by parts. 

\subsection{$AAAc$-vertex}
The $AAAc$-vertex receives contribution from \re{bG1} and the term 
originating from the second iteration of \re{bG0} \ie
\beq\nome{bG2}
(-)^{\de_J}\int_{q'q''} 
\bigg(D_{\L_0\infty}(q'')\,S_{BRS}^{int}(-q'',-p)\bigg)_{HJ}
\bigg(D_{\L_0\infty}(q')\,S_{BRS}^{int}(-q',q'')\bigg)_{LH}
\bigg(S_{BRS}^{int}(-q,q')\bigg)_{IL}\,.
\eeq
The corresponding graphs are shown in figs.~5-7 and the
various contributions are computed in Appendix A. After performing the
trace over the gauge indices this vertex can be written as
$$
\frac{1}{3!}\int_{p_1p_2p_3} \Biggl(A_{\m_1}(p_1)\cdot  A_{\m_2}(p_2)\Biggr)
\Biggl( A_{\m_3}(p_3)\cdot c(-p_1-p_2-p_3)\Biggr) \;
\hat{\D}_{\G\,\m_1\m_2\m_3}^{(AAAc)}(p_1,p_2,p_3;\L_0)\,.
$$
In the $\UV$ limit one obtains
\beeq\nome{d711}
&&\hat{\D}_{\G\,\m_1\m_2\m_3}^{(AAAc)}(p_1,p_2,p_3;\L_0)=
-3i[\hat\de_7 (p_{1\m_1} g_{\m_2\m_3}+
p_{2\m_2} g_{\m_1\m_3})
+  2\hat\de_8 p_{3\m_3} g_{\m_1\m_2}
\nonumber\\
&&
+ \hat\de_9 (p_{1\m_2} g_{\m_1\m_3}+p_{2\m_1} g_{\m_2\m_3})
+ \hat\de_{10}( p_{1\m_3}+p_{2\m_3}) g_{\m_1\m_2}
+  \hat\de_{11} (p_{3\m_2} g_{\m_1\m_3}+p_{3\m_1} g_{\m_2\m_3}
)]\nonumber\\
&&
\phantom{\hat{\D}_{\G\,\m_1\m_2\m_3}^{(AAAc)}(p_1,p_2,p_3;\L_0}
\,+\, {\cal O}(P^4/\L_0^2)\,,
\eeeq
where 
\beq\nome{dh7}
\hat\de_7=\hat\de_9=\hat\de_{11}
=-g^3\int_q\frac{(1-K)}{6q^4}[31K^3+K^2(124K'q^2-47)+K(16-111K'q^2)+14K'q^2]
\eeq
and
\beq\nome{dh8}
\hat\de_8=\frac12\hat\de_{10}
=-g^3\int_q\frac{(1-K)}{6q^4}[23K^3+K^2(56K'q^2-25)+K(2-48K'q^2)+K'q^2]
\,.
\eeq
By using 
the fine-tuning equation \re{fintun}, together with the results
\re{de} and \re{sig3}, 
the couplings $\s_5$ and $\s_6$
are given by
$$
\s_5^{(1)}=-ig^2\biggl(\frac{3r}{2}+\int_q
\frac{K(1-K)(23K^2-12K-11)-10K'^2q^4}{6q^4}\biggr)\,,
$$
$$
\s_6^{(1)}=-ig^2\,\biggl(\frac{2r}{3}+\int_q \frac{(1-K)(9K^3+12K^2+ 3 K)}{q^4}
\biggr)
$$
and 
their values for different cutoff functions are given in table~3.
\begin{table}[htbp]
\begin{center}
\begin{tabular}{|c |c | c| c |} \hline
                   &   &  &      \\
$\Kiu(p)$ & 
$e^{-p^2/\L_0^2}$&
$\frac{\L_0^4}{(p^2+\L_0^2)^2} $& 
$\frac{\L_0^{2n}}{p^{2n}+\L_0^{2n}} $
\\
 & & & \\  \hline
 & & &  \\
$\s_5$ &  
-$\frac{ig^2r(13+94\ln2-70\ln3)}{12}$&
$\frac{47ig^2r}{630}$ &
$\frac{ig^2(5n^2-27n+28)}{18n}$ 
\\
 & & & \\  \hline
 & & &  \\
$\s_6$ &  
-$\frac{ig^2r(2-63\ln{3}+99\ln{2})}{3}$&
-$\frac{58ig^2r}{105}$&
-$\frac{2ig^2}3$
\\
  &  &  &   \\
\hline
\end{tabular} 
\end{center}
\caption[Table 3] {\small{One-loop values of the couplings $\s_5$ and $\s_6$ 
for three different cutoff functions ($r=\frac{i}{16\pi^2}$ and $n\ge 1$).}}
\end{table}

\subsection{$wcc$-vertex}
The $wcc$-vertex is obtained inserting \re{bG1} in \re{dgh1} and is given 
by the three graphs depicted in Fig.~3.  
Using the vertices of $S_{BRS}$ one gets 
\begin{figure}[htbp]
\begin{center}
\epsfysize=4cm
\epsfbox{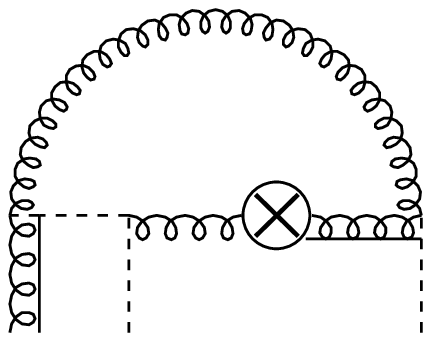}
\epsfysize=4cm
\epsfbox{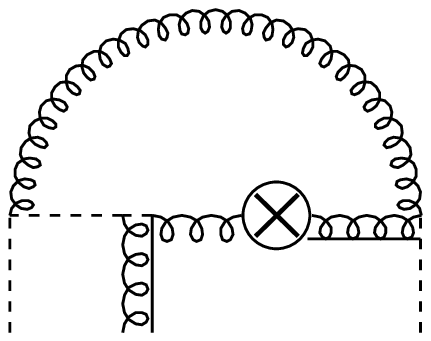}
\epsfysize=4cm
\epsfbox{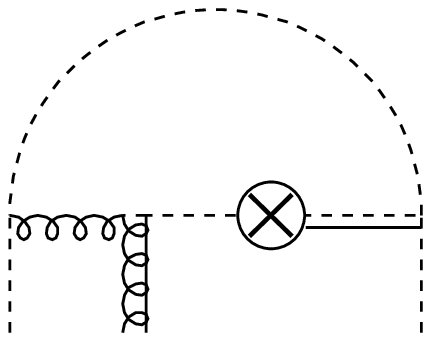}
\end{center}
\caption{\small{Graphical contribution to the $w$-$c$-$c$-vertex of $\DGh$.
}}
\end{figure}
$$
\frac12\int_{p_1p_2} w_\m(-p_1-p_2)
\cdot c(p_1)\wedge c(p_2) \;
\hat{\D}_{\G\,\m}^{(wcc)}(p_1,p_2;\L_0)
$$
with
\beeq
&&\hat{\D}_{\G\,\m}^{(wcc)}(p_1,p_2;\L_0)\,
\nonumber\\&&
\;\;\;\;
=g\,\int_q\Biggl[(q-p_1-p_2)_\m
\frac{\Kin(q)\Kin(q-p_1-p_2)\Kiu(q-p_2)}{q^2 (q-p_1-p_2)^2}\,
\nonumber\\&&
\nonumber\\&&
\;\;\;\;
-\,q_\m\frac{\Kin(q)\Kin(q+p_2)[\Kiu(q+p_1+p_2)+\Kiu(q-p_1)]}{q^2 (q+p_2)^2}\,
\nonumber\\&&
\nonumber\\&&
\;\;\;\;
+(1\leftrightarrow 2)
\Biggr]
\,.
\nonumber
\eeeq

In the $\UV$ limit one obtains 
$$
\hat{\D}_{\G\,\m}^{(wcc)}(p_1,p_2;\L_0)\,
= \,i(p_1+p_2)_\m\,[\hat\de_5\,+\, {\cal O}(P^2/\L_0^2)]\,,
$$
where 
$$
\hat\de_5=g\int_q\frac{(-1+4K-3K^2)K'}{q^2}=0
$$
as integration by parts can show. Thus from \re{de}
the one-loop value of the coupling $\s_4$ is given by \re{sig3}.

\subsection{$wccA$-vertex}
The fine-tuning equation for this vertex is directly obtained form the 
the relation between $\de_5$ and $\de_6$ in \re{de}, thus this vertex
must vanish.

The graphs which contribute to this vertex are obtained 
inserting \re{bG2} in  \re{dgh1} and are given in fig.~4. 
\begin{figure}[htbp]
\begin{center}
\epsfysize=4cm
\epsfbox{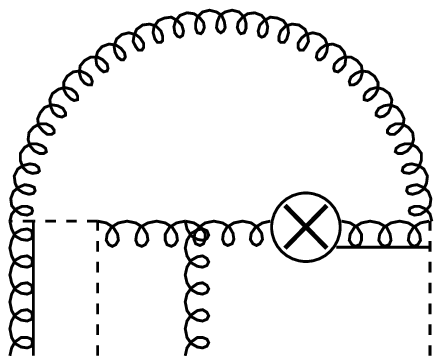}
\epsfysize=4cm
\epsfbox{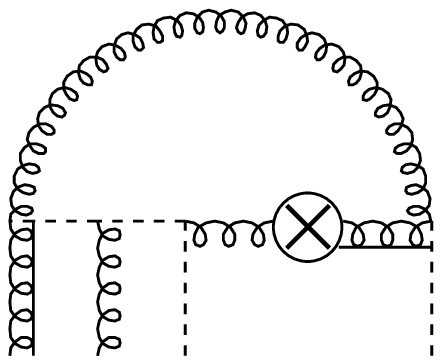}
\epsfysize=4cm
\epsfbox{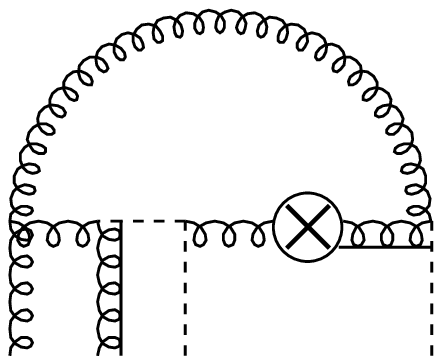}
\epsfysize=4cm
\epsfbox{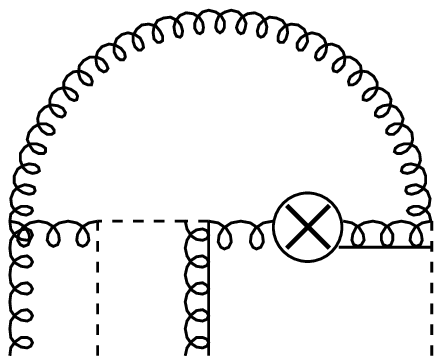}
\epsfysize=4cm
\epsfbox{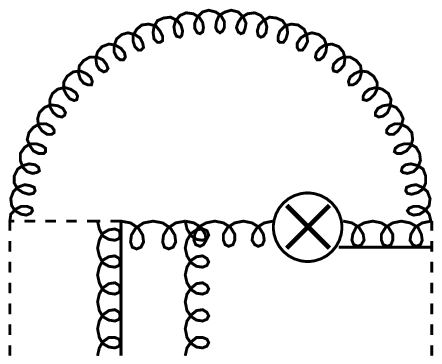}
\epsfysize=4cm
\epsfbox{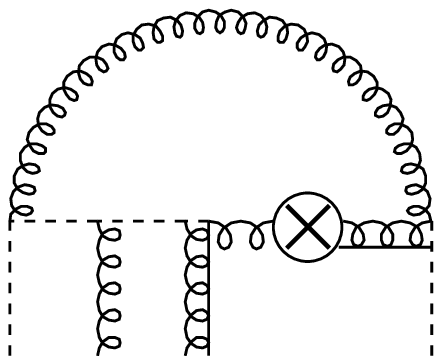}
\epsfysize=4cm
\epsfbox{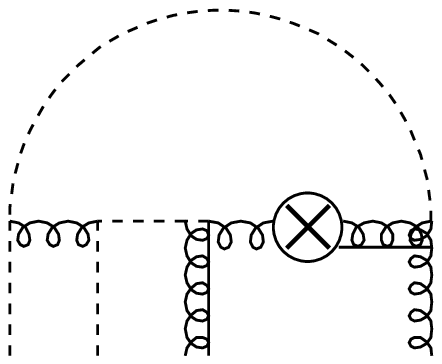}
\epsfysize=4cm
\epsfbox{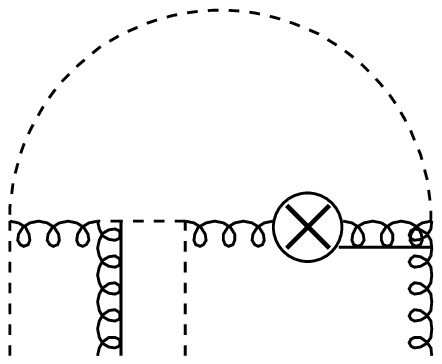}
\epsfysize=4cm
\epsfbox{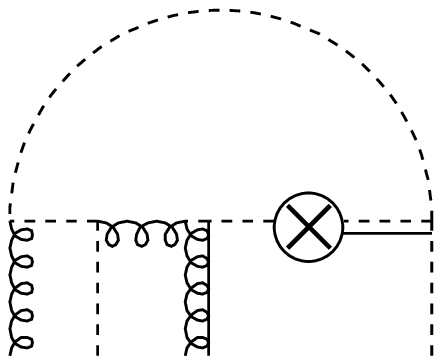}
\epsfysize=4cm
\epsfbox{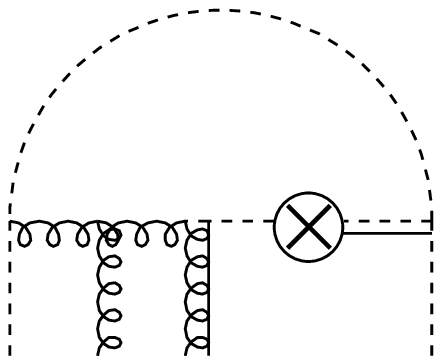}
\epsfysize=4cm
\epsfbox{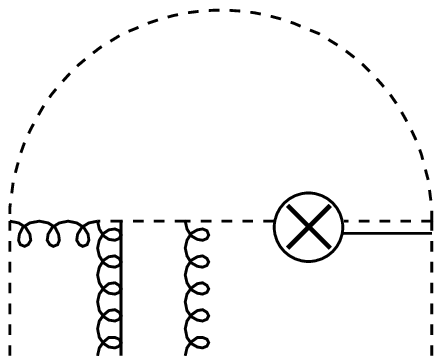}
\end{center}
\caption{\small{Graphical contribution to the $w$-$c$-$c$-$A$-vertex of $\DGh$.
}}
\end{figure}
All the terms can be collected in the following monomial
$$
\int_{p_1p_2p_3}(A_\m(p_3)\cdot c(p_2) \,)( w_\n(-p_1-p_2-p_3)\cdot c(p_1)\,)
\hat{\D}_{\G\,\m\n}^{(Accw)}(p_3,p_2,p_1,p_3;\L_0)
\,.
$$
The complete expression of 
$\hat{\D}_{\G\,\m\n}^{(Accw)}$ is not needed since 
$$
\hat{\D}_{\G\,\m\n}^{(Accw)}(p_3,p_2,p_1,p_3;\L_0)=g_{\m\n}\hat\de_6\
+\, {\cal O}(P^2/\L_0^2)
$$
and therefore in the $\UV$ limit one can set to zero all the external momenta.
In this limit one obtains
$$
g_{\m\n}\hat\de_6
=g^2\int_q\frac{(1-K)^3K}{q^6}
[-q^2 g_{\m\n}+q_\m q_\n+q_\r t_{\r\n\m}(q,0)]
\,,
$$
which vanishes.

\section{Conclusions}
The computation presented in this paper has shown that the requirement
that the physical effective action satisfies the ST identity
translates into a fine-tuning equation which
allows to determine the couplings of the relevant, non-invariant
interactions in the bare action $\Pi(\L_0)$. 
Once the field normalization and the gauge
coupling are fixed at the physical point $\L=0$, the problem of
assigning the boundary conditions of the RG flow is finally solved.
We explicitly compute the non-invariant couplings at one-loop, they
turns out to be finite.
This procedure systematically generalizes to higher orders.  There are
actually two crucial points which deserve mentioning.  First of all
one must discuss the locality of $\DGh$. At one-loop order this has
been explicitly shown in section 4 and is ensured by the fact that at
tree level and at the UV scale the functional 
$(\frac{\de^2\G}{\de\Phi\de\Phi})^{-1}$ 
is given by either a relevant
vertex or a sequence of relevant vertices joint by propagators with a
cutoff function $\Kin(q+P)$, where $P$ is a combination of external
momenta.  Since the integral in \re{dgh} is damped by these cutoff
functions, only the contributions with a restricted number of
propagators survive in the $\UV$ limit. One can infer from power
counting that they are of the relevant type.  A similar argument holds
for the possible non-local contributions coming from $\Kiu(p)$.
Nevertheless at higher loops one could
worry about non-local terms which may arise from the full propagators, 
$\Pi(\L_0)$
and the proper vertices generated in the expansion of
$(\frac{\de^2\G}{\de\Phi\de\Phi})^{-1}$.  
With a little thought, it is easy to realize that
the terms of this expansion which survive in the $\UV$ limit have less
than three propagators, \ie one more than in the one-loop case.  This
related to the fact that at higher loops the $w$-$c$-vertex in
$\Pi(\L_0)$ contributes to \re{dgh} and therefore all the fields of
$\DGh$ originate from the inversion of $\frac{\de^2\G}{\de\Phi\de\Phi}$
(recall that the relevant part of the
functional $\DG$ contains at most four fields).  For the same reason,
irrelevant proper vertices with at most six legs have to be taken 
into account.
By dimensional analysis non-local parts of these vertices are
proportional to negative powers of $\L_0$, yet they can contribute to
the relevant part of $\DGh$ since in \re{dgh} the loop momentum $q$ is
of the order of $\L_0$.  As a consequence higher derivative
interactions (\ie vertices with higher powers of the momenta, but a
restricted number of fields), are also involved in the
fine-tuning. In particular one has to consider irrelevant interactions
involving the BRS sources.  Therefore at higher loops the fine-tuning
procedure is rather cumbersome but well defined since the irrelevant
vertices are completely determined by the flow equation \re{eveq}
while the relevant ones are given by the fine-tuning at the lower
loop orders.

Another issue to discuss is the finiteness of the non-invariant
couplings at higher loops.  The presence in \re{dgh} of the cutoff
functions $\Kiu(q+P_i)$ and $\Kin(q+P_j)$, 
having almost non-intersecting
supports, makes the $q$-integration in \re{dgh} finite.  Actually the
cutoff function $\Kiu(q)$ must fall off more rapidly than any negative
power of $q$ for $q^2>\L_0^2$, in order to compensate the powers of
$q$ in the irrelevant vertices. 
Divergent contributions to the non-invariant couplings $\s_i$ can be
found only when we consider in \re{dgh} the relevant vertices $z_i(\L_0)$,
but they can be re-absorbed in field and gauge coupling redefinition.

Clearly, more elegant formulations exist for a pure gauge theory, such
as dimensional regularization with minimal subtraction, which
preserves BRS invariance and thus avoids the fine-tuning.  However the
RG formulation is general and can be applied to chiral gauge theories
without anomalies along the same lines. In this case all the
regularization procedures break the gauge symmetry and the fine-tuning
is unavoidable. 
Moreover in the dimensional regularization with
minimal subtraction scheme the breaking of the symmetry requires 
the introduction of all the possible non-invariant counterterms~\footnote{
Recently \cite{recent} these counterterms have been systematically 
computed at one-loop oder in the dimensional regularization scheme.} 
while in the RG approach only
interactions which are invariant under global chiral transformation
need to be considered in the fine-tuning procedure \cite{BV}.

As a final remark we should mention that it has been suggested 
\cite{MH} to use gauge invariant variables, such as the Wilson loop, to 
overcome the difficulty of the fine-tuning. 
However is not obvious how  this procedure can be extended to
chiral gauge theories.

\vspace{1cm}\noindent{\large\bf Acknowledgements}
We have greatly benefited  from discussions with G. Marchesini and F. Vian.

\section*{Appendix A}
In this appendix we compute the various contributions to the $AAAc$-vertex.
We first consider  in \re{dgh1} the terms with $\g_i=w_\m$.
In this case both the contributions to the functional $\bG^{(0)}$ 
given in \re{bG1} and \re{bG2} 
are needed to compute this vertex. 
The former involves the four-vector vertex given in \re{ex} and 
the corresponding graphs are shown in Fig.~5.
\begin{figure}[htbp]
\begin{center}
\epsfysize=4cm
\epsfbox{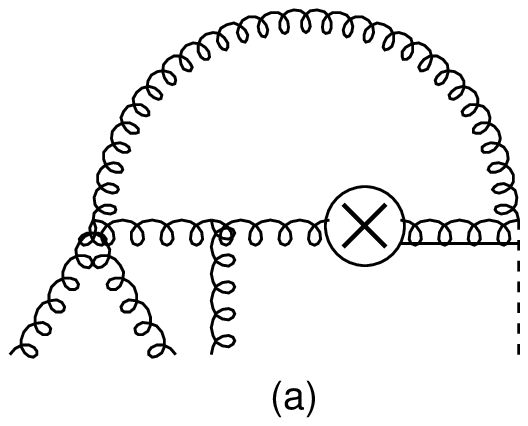}
\epsfysize=4cm
\epsfbox{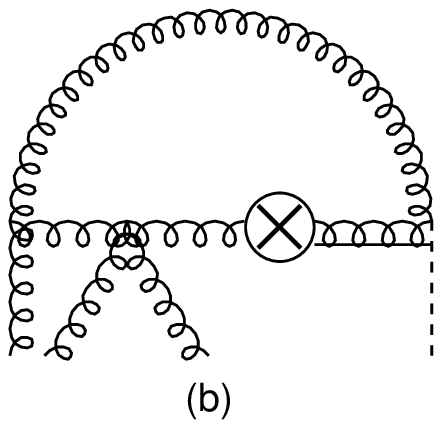}
\epsfysize=4cm
\epsfbox{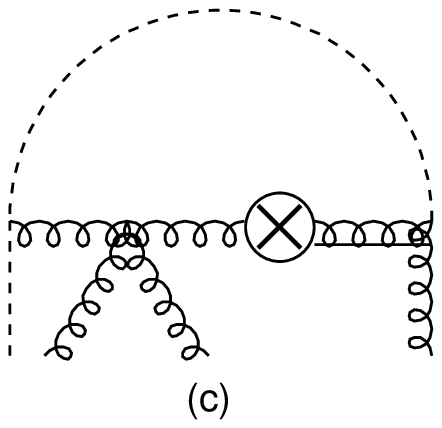}
\end{center}
\caption{\small{Graphical contribution to the $A$-$A$-$A$-$c$-vertex of $\DGh$.
}}
\end{figure}
\newline
The contribution to $\hat{\D}_{\G\,\m_1\m_2\m_3}^{(AAAc)}(p_1,p_2,p_3;\L_0)$
from the the graph in fig.~5a is
(recall that $p_1$, $p_2$ and $p_3$
are the momenta of the vector fields and that one has to consider the 
three permutations obtained by choosing the two external vectors which enters 
in the four-vector vertex) 
\beeq&&
3ig^3\int_q
\frac{\Kin(q)}{q^2}\Kiu(q+p_1+p_2+p_3)
\Bigg\{2g_{\m_1\m_2} t_{\m\m_3\m}(q+p_{12},p_3)
\frac{\Kin(q+p_{12})}{(q+p_{12})^2}
\nonumber\\
&&
\phantom{3\int_q}
+
g_{\m_1\m_3} t_{\m\m_2\m}(q+p_{13},p_2)
\frac{\Kin(q+p_{13})}{(q+p_{13})^2}
+g_{\m_2\m_3}t_{\m\m_1\m}(q+p_{23},p_1)
\frac{\Kin(q+p_{23})}{(q+p_{23})^2}
\nonumber\\
&&
\phantom{3\int_q}
+[t_{\m_1\m_2\m_3}(q+p_{13},p_2)
-2t_{\m_3\m_2\m_1}(q+p_{13},p_2)]
\frac{\Kin(q+p_{13})}{(q+p_{13})^2}
\nonumber\\
&&
\phantom{3\int_q}
+[t_{\m_2\m_1\m_3}(q+p_{23},p_1)-2t_{\m_3\m_1\m_2}(q+p_{23},p_1)]
\frac{\Kin(q+p_{23})}{(q+p_{23})^2}
\nonumber\\
&&
\phantom{3\int_q}
-[t_{\m_1\m_3\m_2}(q+p_{12},p_3)+t_{\m_2\m_3\m_1}(q+p_{12},p_3)]
\frac{\Kin(q+p_{12})}{(q+p_{12})^2}
\Bigg\}\,.
\nonumber
\eeeq
For $\UV$ this graph gives the following contribution 
to the various coefficients 
$\hat\de_i$ in \re{d711}
\beeq&&
\hat\de_7^{(5a)}=
\int_q\frac{(1-K)(7K^2-7K'q^2-K(7-10K'q^2))}{2q^4}
\nonumber\\&&
\hat\de_8^{(5a)}=
\int_q\frac{(1-K)(5K^2-5K'q^2-K(5-6K'q^2))}{2q^4}
\nonumber\\&&
\hat\de_9^{(5a)}=
-\int_q\frac{(1-K)(2K^2+7K'q^2-K(2+13K'q^2))}{2q^4}
\nonumber\\&&
\hat\de_{10}^{(5a)}=
\int_q\frac{(1-K)(19K^2-10K'q^2-19K(1-K'q^2))}{2q^4}
\nonumber\\&&
\hat\de_{11}^{(5a)}=
\int_q\frac{(1-K)(7K^2-7K'q^2-K(7-12K'q^2))}{2q^4}\,.
\nonumber
\eeeq

The graph in Fig.~5b gives 
\beeq&&
3ig^3\int_q
\frac{\Kin(q)}{q^2}\Kiu(q+p_1+p_2+p_3)
\Bigg\{2g_{\m_1\m_2} t_{\m\m_3\m}(q,p_3)
\frac{\Kin(q+p_3)}{(q+p_{3})^2}
\nonumber\\
&&
\phantom{3\int_q}
+
g_{\m_1\m_3} t_{\m\m_2\m}(q,p_2)
\frac{\Kin(q+p_{2})}{(q+p_{2})^2}
+g_{\m_2\m_3}t_{\m\m_1\m}(q,p_1)
\frac{\Kin(q+p_{1})}{(q+p_{1})^2}
\nonumber\\
&&
\phantom{3\int_q}
-[2t_{\m_1\m_2\m_3}(q,p_2)
-t_{\m_3\m_2\m_1}(q,p_2)]
\frac{\Kin(q+p_{2})}{(q+p_{2})^2}
\nonumber\\
&&
\phantom{3\int_q}
-[2t_{\m_2\m_1\m_3}(q,p_1)-t_{\m_3\m_1\m_2}(q,p_1)]
\frac{\Kin(q+p_{1})}{(q+p_{1})^2}
\nonumber\\
&&
\phantom{3\int_q}
-[t_{\m_1\m_3\m_2}(q,p_3)+t_{\m_2\m_3\m_1}(q,p_3)]
\frac{\Kin(q+p_{3})}{(q+p_{3})^2}
\Bigg\}\,.
\nonumber
\eeeq
In the $\UV$ limit the various coefficients 
$\hat\de_i$  for this term are given by
\beeq&&
\hat\de_7^{(5b)}=
\int_q\frac{(1-K)(11K-7)K'}{2q^4}
\nonumber\\&&
\hat\de_8^{(5b)}=
\int_q\frac{(1-K)(9K-5)K'}{2q^4}
\nonumber\\&&
\hat\de_9^{(5b)}=
\int_q\frac{(1-K)(9K^2-9K-7K'q^2+8KK'q^2)}{2q^4}
\nonumber\\&&
\hat\de_{10}^{(5b)}=
-\int_q\frac{(1-K)(9K^2-9K+10K'q^2-11KK'q^2)}{2q^4}
\nonumber\\&&
\hat\de_{11}^{(5b)}=
\int_q\frac{(1-K)(9K-7)K'}{2q^4}\,.
\nonumber
\eeeq

Finally from Fig.~5c we get 
\beeq
&&
-3ig^3\int_q
\frac{\Kin(q)}{q^2}\Kiu(q+p_1+p_2+p_3)
\Bigg\{
\frac{\Kin(q-p_3)}{(q-p_3)^2}
[2g_{\m_1\m_2} q_{\m_3}-g_{\m_2\m_3} q_{\m_1}-g_{\m_1\m_3} q_{\m_2}]
\nonumber\\
&&
\phantom{-3ig^2}
+\frac{\Kin(q-p_2)}{(q-p_2)^2}
[g_{\m_1\m_3} q_{\m_2}+g_{\m_1\m_2} q_{\m_3}-2g_{\m_2\m_3} q_{\m_1}]
\nonumber\\
&&
\phantom{-3ig^2}+
\frac{\Kin(q-p_1)}{(q-p_1)^2}
[g_{\m_2\m_3} q_{\m_1}+g_{\m_1\m_2} q_{\m_3}-2g_{\m_1\m_3} q_{\m_2}]
\Bigg\}\,.
\nonumber
\eeeq
The $\hat\de_i$ for this graph are 
\beeq&&
\hat\de_7^{(5c)}=
-\int_q\frac{(1-K)(2K^2-2K-K'q^2-KK'q^2)}{2q^4}
\nonumber\\&&
\hat\de_8^{(5c)}=-\hat\de_{11}^{(5c)}=
\int_q\frac{(1-K)(2K^2-2K+K'q^2-3KK'q^2)}{2q^4}
\nonumber\\&&
\hat\de_9^{(5c)}=
-\int_q\frac{(1-K)(K^2-K+K'q^2-2KK'q^2)}{q^4}
\nonumber\\&&
\hat\de_{10}^{(5c)}=
\int_q\frac{(1-K)(4K^2-4K+K'q^2-5KK'q^2)}{2q^4}\,.
\nonumber
\eeeq

All the remaining contributions to
$\hat{\D}_{\G\,\m_1\m_2\m_3}^{(AAAc)}(p_1,p_2,p_3;\L_0)$ are generated
from the part of the functional $\bG^{(0)}$ given in \re{bG2}.  For
$\g_i=w_\m$ there are four different terms according to the
position of the ghost field in \re{bG2} which are depicted in
Fig.~6.  
\begin{figure}[htbp]
\begin{center}
\epsfysize=4cm
\epsfbox{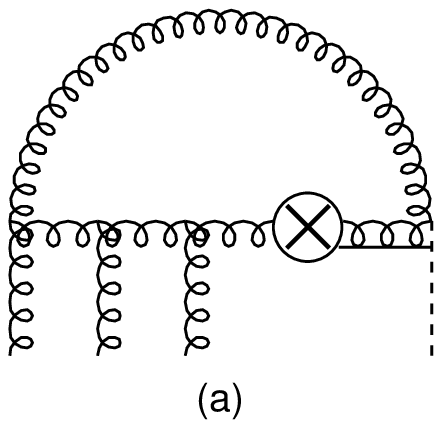}
\epsfysize=4cm
\epsfbox{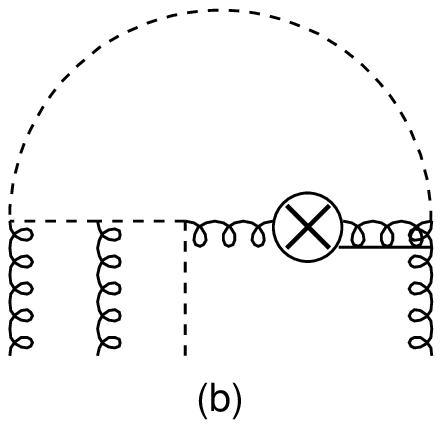}
\epsfysize=4cm
\epsfbox{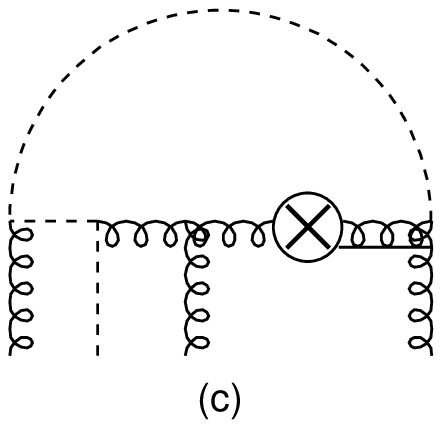}
\epsfysize=4cm
\epsfbox{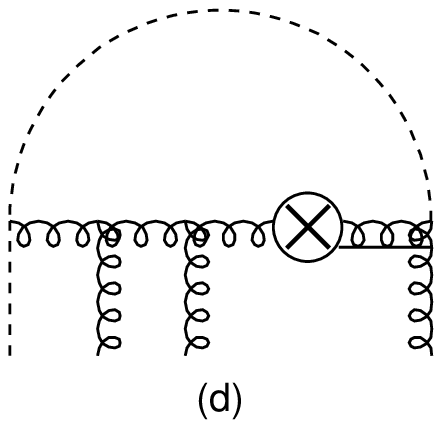}
\end{center}
\caption{\small{Graphical contribution to the $A$-$A$-$A$-$c$ vertex of $\DGh$.
}}
\end{figure}
The contribution from the graph in Fig.~6a is
\beeq
&&-ig^3
\,\int_q\frac{\Kin(q)}{q^2}\Kiu(q+p_1+p_2+p_3)
\nonumber\\
&&\phantom{-ig^3}\times
\Bigg\{
t_{\m\m_1\n}(q,p_1)\,t_{\n\m_2\r}(q+p_1,p_2)\,\,t_{\r\m_3\m}(q+p_{12},p_3)
\frac{\Kin(q+p_1)\Kin(q+p_{12})}
{(q+p_1)^2(q+p_{12})^2}
\nonumber\\
&&\phantom{-ig^3}
+
t_{\m\m_3\n}(q,p_3)\,t_{\n\m_2\r}(q+p_3,p_2)\,\,t_{\r\m_1\m}(q+p_{23},p_1)
\frac{\Kin(q+p_3)\Kin(q+p_{23})}
{(q+p_3)^2(q+p_{23})^2}
\Bigg\}
\nonumber\\
&&\phantom{-ig^3}\
+{\mbox{permutations}}
\,,
\nonumber
\eeeq
where the permutations are performed among the vector fields.
For $\UV$ the contribution of this graph to the various coefficients 
$\hat\de_i$ in \re{d711} is 
\beeq&&
\hat\de_7^{(6a)}=2\hat\de_8^{(6a)}=
\hat\de_9^{(6a)}=\hat\de_{10}^{(6a)}=\hat\de_{11}^{(6a)}=
\nonumber\\&&
-5\int_q\frac{(1-K)^2(K^2-2K'q^2-K(1-4K'q^2))}{q^4}
\,.
\nonumber
\eeeq

The graph in Fig.~6b gives 
\beeq
&&ig^3
\,\int_q\frac{\Kin(q)\Kin(q+p_1)}{q^2(q+p_1)^2}\,q_{\m_1}
\Bigg\{
(q+p_1)_{\m_2}(q+p_{12})_{\m_3}
\frac{\Kin(q+p_{12})}
{(q+p_{12})^2}\Kiu(q-p_3)
\nonumber\\
&&\phantom{ig^3\,\int_q}
+
(q+p_1)_{\m_3}(q+p_{13})_{\m_2}
\frac{\Kin(q+p_{13})}
{(q+p_{13})^2}\Kiu(q-p_2)
\Bigg\}\;\;
+{\mbox{permutations}}
\,.
\nonumber
\eeeq
The $\hat\de_i$ for this graph are
\beeq&&
\hat\de_7^{(6b)}=
\int_q\frac{(1-K)^2(5K^2-K'q^2-K(5+4K'q^2))}{12q^4}
\nonumber\\&&
\hat\de_8^{(6b)}=
\int_q\frac{(1-K)^2(K^2-K'q^2-K)}{12q^4}
\nonumber\\&&
\hat\de_9^{(6b)}=
-\int_q\frac{(1-K)^2(K^2+K'q^2-K(1-4K'q^2))}{12q^4}
\nonumber\\&&
\hat\de_{10}^{(6b)}=
-\int_q\frac{(1-K)^2(4K^2+K'q^2-K(1-4K'q^2))}{12q^4}
\nonumber\\&&
\hat\de_{11}^{(6b)}=
-\int_q\frac{(1-K)^2(K^2+2K'q^2-K)}{12q^4}\,.
\nonumber
\eeeq

From Fig.~6c  we get the contribution 
\beeq
&&-ig^3
\,\int_q 
\Bigg\{
q_{\m_3}(q+p_3)_\m t_{\m\m_2\m_1}(q-p_{12},p_2)
\frac{\Kin(q+p_3)\Kin(q-p_{12})}{(q+p_3)^2(q-p_{12})^2}
\Kiu(q-p_1)
\nonumber\\
&&\phantom{-ig^3}
+
q_{\m_1}(q+p_1)_\m t_{\m\m_3\m_2}(q-p_{23},p_3)
\frac{\Kin(q+p_1)\Kin(q-p_{23})}{(q+p_1)^2(q-p_{23})^2}
\Kiu(q-p_2)
\Bigg\}\;\;
\nonumber\\
&&\phantom{-ig^3}
\times
\frac{\Kin(q)}{q^2}
+{\mbox{permutations}}
\nonumber
\eeeq
which gives 
\beeq&&
\hat\de_7^{(6c)}=
-\int_q\frac{(1-K)^2(4K^2+K'q^2-K(4+5K'q^2))}{6q^4}
\nonumber\\&&
\hat\de_8^{(6c)}=-\hat\de_{11}^{(6c)}=
\int_q\frac{(1-K)^2KK'}{2q^4}
\nonumber\\&&
\hat\de_9^{(6c)}=
-\int_q\frac{(1-K)^2(5K^2-4K'q^2-K(5-8K'q^2))}{12q^4}
\nonumber\\&&
\hat\de_{10}^{(6c)}=
\int_q\frac{(1-K)^2(19K^2+4K'q^2-K(19+14K'q^2))}{12q^4}
\,.
\nonumber
\eeeq

From Fig.~6d one has 
\beeq
&&ig^3
\,\int_q q_{\m}
\Bigg\{
t_{\m\m_1\n}(q-p_1-p_2-p_3,p_1)
t_{\n\m_2\m_3}(q-p_{23},p_2)
\frac{\Kin(q-p_{23})}
{(q-p_{23})^2}\Kiu(q-p_3)
\nonumber\\
&&\phantom{ig^3}
+
t_{\m\m_3\n}(q-p_1-p_2-p_3,p_3)
t_{\n\m_2\m_1}(q-p_{12},p_2)
\frac{\Kin(q-p_{12})}
{(q-p_{12})^2}\Kiu(q-p_1)
\Bigg\}
\nonumber\\
&&\phantom{ig^3}
\times
\frac{\Kin(q)\Kin(q-p_1-p_2-p_3)}{q^2(q-p_1-p_2-p_3)^2}
+{\mbox{permutations}}
\nonumber
\eeeq
and finds 
\beeq&&
\hat\de_7^{(6d)}=
\int_q\frac{(1-K)^2(5K^2-7K'q^2-K(5+22K'q^2))}{12q^4}
\nonumber\\&&
\hat\de_8^{(6d)}=
-\int_q\frac{(1-K)^2(15K^2+7K'q^2-K(15+22K'q^2))}{12q^4}
\nonumber\\&&
\hat\de_9^{(6d)}=
\int_q\frac{(1-K)^2(2K^2+11K'q^2-2K(1+14K'q^2))}{12q^4}
\nonumber\\&&
\hat\de_{10}^{(6d)}=
-\int_q\frac{(1-K)^2(49K^2+K'q^2-K(49+26K'q^2))}{12q^4}
\nonumber\\&&
\hat\de_{11}^{(6d)}=
-\int_q\frac{(1-K)^2(3K^2-10K'q^2-K(3-34K'q^2))}{12q^4}\,.
\nonumber
\eeeq

Finally there is a contribution to 
$\hat{\D}_{\G\,\m_1\m_2\m_3}^{(AAAc)}(p_1,p_2,p_3;\L_0)$
for  $\g_i=v$ which is  depicted in Fig.~7 and gives   
\begin{figure}[htbp]
\epsfysize=4cm
\begin{center}
\epsfbox{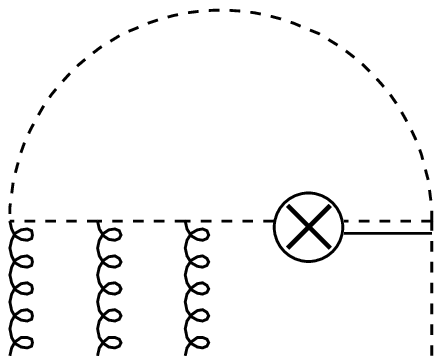}
\end{center}
\caption{\small{Graphical contribution to the $A$-$A$-$A$-$c$ vertex of $\DGh$.
}}
\end{figure}
\newline
\beeq
&&-ig^3
\,\int_q\frac{\Kin(q)}{q^2}\Kiu(q+p_1+p_2+p_3)
\nonumber\\
&&\phantom{-ig^3}\times
\Bigg\{
q_{\m_1}(q+p_1)_{\m_2}(q+p_{12})_{\m_3}
\frac{\Kin(q+p_1)\Kin(q+p_{12})}
{(q+p_1)^2(q+p_{12})^2}
\nonumber\\
&&\phantom{-ig^3}
+
q_{\m_3}(q+p_3)_{\m_2}(q+p_{23})_{\m_1}
\frac{\Kin(q+p_3)\Kin(q+p_{23})}
{(q+p_3)^2(q+p_{23})^2}
\Bigg\}
\nonumber\\
&&\phantom{-ig^3}\
+{\mbox{permutations}}
\,.
\nonumber
\eeeq
The contribution to the $\hat\de_i$ of this graph
$$
\hat\de_7^{(7)}=2\hat\de_8^{(7)}=
\hat\de_9^{(7)}=\hat\de_{10}^{(7)}=\hat\de_{11}^{(7)}=
-\int_q\frac{(1-K)^2(K^2+K'q^2-K(1+2K'q^2))}{3q^4}\,.
$$

After putting together all these results one obtains \re{dh7} and \re{dh8}.

\end{document}